# Analysis of the LockBit 3.0 and its infiltration into Advanced's infrastructure crippling NHS services

Oladipupo Akinyemi
Faculty of Engineering and Environment
Northumbria University
London, United Kingdom
oladipupo.akinyemi@nortumbria.ac.uk

Rejwan Sulaiman
Faculty of Engineering and Environment
Northumbria University
London, United Kingdom
rejwan.sulaiman@northumbria.ac.uk

Nasr Abosata
Faculty of Engineering and Environment
Northumbria University
London, United Kingdom
nasr.abosata@northumbria.ac.uk

*Abstract*—The LockBit 3.0 ransomware variant is arguably the most threatening of malware in recent times. With no regard for a victim's industry, the ransomware has undergone several evolutions to arrive at an adaptable and evasive variant which has been a menace to governments and organisations, recently infiltrating Advanced Computer Software group. Previous Lockbit studies mostly concentrated on measuring the impact of the ransomware attack, prevention, encryption detection, decryption, or data recovery, thereby providing little or no benefit to the less tech savvy populace as a detailed breakdown of the mode of attack is rarely examined. This article analyses the LockBit 3.0 attack techniques with a contextual illustration of the attack on Advanced Computer Software group. With the NHS being a major client of the organisation, and its services alongside 15 other clients being crippled for hours during the attack, attention is drawn to how dreadful such disruption may be in critical organisations. We observed that the upgrade of Lockbit based on releasing newer versions is in a bid to continuously ensure the malware's efficiency – a virtue that keeps it at the zenith – by staying ahead of improved defenses. Our study highlights social engineering as a vibrant portal to Lockbit's maliciousness and indicates an investment in detection systems may profit more than in prevention systems. Therefore, further research should consider improving detection systems against Lockbit 3.0.

## I. INTRODUCTION

Since its inception, LockBit a ransomware-as-a-service (RaaS) has proven to be a large-scale threat. Discovered in 2019 and initially dubbed as the ".abcd virus" - coined from the victim's encrypted file extension – [1], the ransomware is connected to a third of ransomware attacks recorded since the second half of 2022 till date [2]. Although Blackberry [3] depicts LockBit targets as small-to-medium sized organisations due to a ridiculously cheap ransom demanded when com- pared to an average ransomware payment, Abrams [4] claims LockBit threat actors demanded a huge 8 million dollars ransom from one of their recent victims. According to Trend Micro [5] which portrays healthcare, education, technology, governments, oil and gas, manufacturing, transportation etc. industries as regular targets, the threat actors have shown they are in for a long haul. Furthermore, Kaspersky [1] claims that the increase in remote working could also be a factor for not ruling out the possibility of this sort of attack across industries and organisations.

Over the years, there have been different versions of LockBit. Likened to an evolutionary variant which developed from the .abcd extension, to the .lockbit extension, and then the LockBit 2.0 version [1]. The latest variant – LockBit 3.0 – which is also known as LockBit Black, is the most adaptable and evasive of the versions [6]. SOCRadar [2] claims that, even though the LockBit 3.0 is known to target Windows, Linux and VMware ESXi servers, new versions capable of targeting macOS, MIPS, ARM, FreeBSD and SPARC servers have been recently identified. It is therefore important that organisations strengthen defences as it is possible that there may be proportionate increase in LockBit attacks as more target devices become penetrable.

Amazingly, the assumption that Cyber Insurance could help mitigate ransomware attacks is somewhat plausible. [7] recalls how a LockBit operator insinuated that an attack on a cyber insured firm guarantees a successful payment. Also, a member of the REvil group once referred to cyber insured companies as "one of the tastiest morsels", and then explained how it was resourceful to hack cyber insurers as reconnaissance before attacking the clients [8]. It is therefore important that organisations become more intentional about dutifully securing their data.

This paper aims to examine the LockBit 3.0 Ransomware with the attack on Advanced software provider which crippled the UK NHS 111 services as a case study. Understanding that Advanced issued a statement confirming that 16 of their Staffplan and Caresys Customer companies were affected by the ransomware attack [9], this article seeks to analyze the attack to proffer futuristic countermeasures. However, it is important to note that conclusions from previous studies addressing encryption, decryption, recovery etc., rarely accounted for the swift transition of Lockbit from one version to another, thereby sometimes rendering their proposed solutions prematurely unhelpful. Therefore, having a broad understanding of how the attack is carried out is likely going to be beneficial as patterns are less susceptible to change. With the aim of the research being learning from the attack to bolster the knowledge required to administer countermeasures; relevant journals, research papers, government publications and online articles pertaining to recent ransomware attacks were reviewed for appropriate insights.

Section 2 provides an analysis of a LockBit 3.0 attack and narrates a possible attack scenario with respect to Advanced's infiltration. Section 3 addresses the general countermeasures for ransomware attacks and further highlights standard recommendations with respect to LockBit 3.0. Section 4 concludes the article by providing insightful inferences established in the writing.

## II. Attack Analysis

Conventionally, ransomware threat actors require access to the victim's infrastructure. In the case of LockBit, initial access could be via Remote Desktop Protocol (RDP) exploitation, abuse of valid accounts, phishing campaigns, public- facing applications exploitation, or drive-by compromise [6]. The attack on Advanced was reported to have been initiated with legitimate third-party credentials [10], suggesting either an abuse of valid accounts or a productive phishing campaign. Furthermore, Advanced stated that in the initial phase of the attack, the attacker moved laterally in the health and care environment and escalated privileges allowing reconnaissance and the deployment of encryption malware.

CyberSecurity & Infrastructure Security Agency [6] claims privilege escalation could entail gathering system information such as hostname, domain information etc., stopping services, executing commands, enabling automatic logon for persistent and privilege escalation, and deleting log files, shadow copies, and files in the recycle bin.

Although, there are several other vulnerabilities the attacker could as well exploit. For example, research by Microsoft showed two vulnerabilities in PaperCut's print management software which when exploited allowed LockBit attackers to install remote management software [11]. Understanding that the PaperCut's software is widely used across industries with no exception to the healthcare, exploitation of these vulnerabilities could have created an avenue for attackers to move laterally through systems, collect information and launch secondary attacks.

The Lateral movement within a target network is eased by utilizing either a predetermined list of credentials hardcoded during the compilation process, or a local account with escalated privileges that has already been compromised. Once compiled, features which may also allow it to spread through Group Policy Objects and PsExec are activated utilizing the Server Message Block (SMB) protocol [6]. The goal of this infiltration is to ensure that recovery is almost impossible without the assistance of the attacker.

Common to ransomware attacks, information theft is considered vital before encryption. Advanced claimed the attackers were able to steal some information before encryption [9]. Newman [12] explains the motive behind such theft is to threaten victims of publicly exposing the information if ransom isn't paid. He further explained that some attackers fancy double extortion, i.e., files are encrypted in two layers, each requiring a ransom for decryption. Trend Micro Research
[5] claims tools like MEGA, FreeFileSync and StealBit – exclusive to LockBit – are used in exfiltrating stolen files.

Considering that Advanced, being a British organisation will most likely have systems language set to English. Lakshmanan [13] claims that the ransomware was programmed to infect only systems configured with languages outside the exclusion list including Romanian (Moldova), Arabic (Syria), and Tatar (Russia). Therefore, the attack program halts when any of these languages is detected.

After the network is readied for LockBit to operate, the ransomware will initiate its spread across any accessible device. LockBit can achieve this with minimal requirements. With a single device that has elevated privileges, it can send out commands to other devices in the network to download and execute the ransomware [1].

The encryption function encrypts all system files, essentially" locking" them, making them inaccessible to the victim, ensuring only a custom decryption key created by LockBit's proprietary decryption tool can create access.

Although Advanced does not publicly provide precise details of the attack, stating that the information would only be provided to customers on request. The usual trajectory for LockBit 3.0 attackers after encryption is to drop ransom notes named "<Ransomware ID >.README.txt", and change host's icon and wallpaper to attacker's customized images [6]. Furthermore, depending on options set at compilation time, LockBit 3.0 may choose to delete itself and/or any update made.

LockBit affiliates use multiple freeware and open-source tools in their attacks. These tools have been used for various activities, including network reconnaissance, remote access and tunnelling, credential dumping, and file exfiltration. Cybersecurity & Infrastructure Security Agency [6] claims the use of PowerShell and Batch scripts is common in most attacks, which concentrate on system discovery, reconnaissance, password and credential hunting, and privilege escalation. Also, signs of professional penetration-testing tools like Metasploit and Cobalt Strike have been identified.

### A. Attack narration

Even though the post-attack information provided by Advanced was superficial, the few details extracted will be used to connect dots and summarize the attack.

- Social engineering was deployed.
- Third-party credentials stolen.
- Stolen credentials used to access management system, evading authentication errors as credential hashes match.

TABLE I
MITRE TACTICS AND TECHNIQUES

| Initial access | Execution | Persistence | Privilege escalation | Defence evasion | Discovery | Lateral movement | Exfiltration | Impact |
|---|---|---|---|---|---|---|---|---|
| T1566 - Phishing | T1204 - User Execution | T1547 - Boot or logon autostart execution | T1134 - Access token manipulation | T1562 - Impair defences | T1083 - File and directory discovery | T1570 - Lateral tool transfer | T1567 - Exfiltration over web service | T1486 - Data encrypted for impact |
| T1078 - Valid accounts | | | | | T1135 - Network Share Discovery | | | T1489 - Service stop |
| | | | | | | | | T1491 - Defacement |

<sup>a</sup>Mapped from Trend Micro and Cybersecurity Infrastructure Security Agency.

- Group policy created to disable security products e.g., windows defender.
- Management system leveraged and LockBit executed via PowerShell Empire.
- Further credential theft using Mimikatz.
- Using the management system, privileges were escalated to enable reconnaissance and facilitate deploying encryption malware.
- Enumeration by Network scanner and port scanner is carried out to locate domain controller or active directory as they are mostly viable for deploying ransomware with network encryption.
- Lateral movement commenced by self-propagation via SMB protocol using obtained credentials and Group policies.
- Exfiltration occurred with some files stolen and uploaded into cloud storage.
- Advanced's health and care environment Systems infected and files encrypted.
- Ransom note created, icons and wallpapers changed to notify victim.

Table 1 gives an insight into the MITRE tactics the Lockbit 3.0 ransomware may have used against Advanced.

## III. COUNTERMEASURES

Arguably one of the biggest threats to individuals and organizations globally, several approaches to detect and prevent ransomware have been identified. However, McIntosh [14] claims the inability of many anti-ransomware studies to account for the evolution of ransomware from executable files encrypting victim files, to the inclusion of fileless command scripts, information exfiltration and human-operated form, could be the reason some recent measures are futile. Although Irwin [15] pins successful ransomware attacks on victim illiteracy, it is important to note that user awareness is only one of the minimum practices required.

There are no elementary solutions to totally protect an organisation from ransomware. Young [16] explains that preventing a ransomware attack requires adopting a layered approach using a defense-in-depth methodology which involves implementing various measures such as providing regular training to users, filtering emails for suspicious content, using virus detection software, configuring firewalls, strengthening edge security, monitoring activities, and employing additional techniques depending on the available budget, resources, and expertise.

Pertaining to the LockBit 3.0, the Cross-Sector Cybersecurity Performance Goals (CPGs) prepared by the CISA and NIST recommend the following set of practices and protections for organizations.

- Establish and enforce a recovery strategy that includes the preservation of multiple copies of sensitive data in a physically isolated and well-protected offline facility.
- Comply with NIST standards regarding password policies.
- Require phishing-resistant multifactor authentication.
- Regularly patch Operating System, software, and firmware.
- Segment networks to curb ransomware spread.
- Log and report all network traffic, including lateral movement activity, to identify, detect, and investigate abnormalities.
- Install antivirus software on all hosts and ensure enabling real time detection and regular updates.
- Observe servers, domain controllers, and active directories for unrecognized accounts.
- Disable unused ports.
- Tag emails received from outside sources with an email banner.
- Disable hyperlinks in emails received.
- Restrict command-line and scripting activities permissions to avoid lateral movement or privilege escalation.
- Ensure backup data is encrypted, contained of all organisations proprietary data and is immutable.

More also, regular testing of existing controls assessing their performance against the MITRE techniques mapped out is recommended.

## IV. Conclusions

In this paper, the recent LockBit variant 3.0 was analyzed relative to the recent attack on Advanced Computer Software group and its customers with one being the UK NHS. The objective was on a premise that a comprehensive understanding of how the Lockbit 3.0 ransomware infiltrated network infrastructures could further aid the prevention of the malware and its likes in the future, by providing insights to its attack pattern, and recommendations for countermeasures. This paper establishes the level of influence LockBit 3.0 has on the ransomware market. It was found that evolution of LockBit over time was to enhance adaptation to its target environment and evade defensive measures, currently making its perpetrators the biggest on the scene. From reports, the paper was able to illustrate from a LockBit attack scenario on Advanced that credentials obtained from social engineering is able to create pathways for an attacker to move laterally within an infrastructure, escalate privileges, exfiltrate and encrypt data. However, it is important to acknowledge the limitations of this study. While the analysis drawn may seem agreeable, the lack of direct intel from Advance computer software group restricted the information gathered for this study to third party study, thereby impeding accuracy. In the end, this study addressed how there is no straightforward countermeasure for ransomware and specifically highlights recommendations to mitigate LockBit 3.0 attacks. Valuable insights which may be resourceful for prevention and detection have been provided from the attack narration and analysis. However, given the rising sophistication of ransomware attacks, it is crucial to regularly evaluate the effectiveness of detection and preventive measures to identify any shortcomings that need to be addressed.

## References

[1] Kaspersky, "LockBit ransomware - What you need to know" Kaspersky.com [Online]. Available: https://www.kaspersky.com/resource-center/threats/lockbit-ransomware [Accessed May 14, 2023].

[2] SOCRadar, "Dark Web Profile: LockBit 3.0 Ransomware." socradar.io, para. 1, Apr. 27, 2023.[Online]. Available: https://socradar.io/dark-web-profile-lockbit-3-0-ransomware [Accessed May 14, 2023].

[3] Blackberry, "What is LockBit Ransomware?" blackberry.com para. 1. [Online]. Available: https://www.blackberry.com/us/en/solutions/endpoint-security/ransomware-protection/lockbit [Accessed. May 10, 2023].

[4] L. Abrams. "LockBit ransomware blames Entrust for Ddos attacks on leak sites." bleepingcomputer.com para. 11, Aug. 22, 2022. [Online]. Available: https://www.bleepingcomputer.com/news/security/lockbit-ransomware-blames-entrust-for-ddos-attacks-on-leak-sites/ [Accessed. May 10, 2023].

[5] Trend Micro Research, "LockBit." trendmicro.com para. 13, Feb. 8, 2022. [Online]. Available: https://www.trendmicro.com/vinfo/us/security/news/ransomware-spotlight/ransomware-spotlight-lockbit [Accessed. May 11, 2023].

[6] Cybersecurity & Infrastructure Security Agency "StopRansomware: LockBit 3.0." cisa.gov para. 2, Mar. 16, 2023. [Online]. Available: https://www.cisa.gov/news-events/cybersecurity-advisories/aa23-075a [Accessed May. 12, 2023].

[7] A. Khodjibaev, D. Korzhevin, and K. McKay, "Interview With a LockBit ransomware Operator New York." blog.talosintelligence.com Feb. 2, 2021. [Online]. Available: https://blog.talosintelligence.com/interview-with-lockbit-ransomware/ [Accessed. May 10, 2023].

[8] G. Mott el al., (2023) "Between a rock and a hard(ening) place: Cyber insurance in the ransomware era," Computer & Security, vol. 128. doi: 10.1016/j.cose.2023.103162.

[9] C. Glover, "LockBit 3.0 used in ransomware attack on Advanced that knocked out NHS 111 services." bleepingcomputer.com para. 1, Oct. 14, 2022. [Online]. Available: https://www.bleepingcomputer.com/news/security/lockbit-ransomware-blames-entrust-for-ddos-attacks-on-leak-sites/ [Accessed May. 14, 2023].

[10] The stack, "Advanced confirms attack was LockBit 3.0 ransomware, legitimate creds used." thestack.technology para. 1, Oct. 13, 2022. [Online]. Available: https://thestack.technology/advanced-data-breach-credentials-ransomware-post-incident-summar/ [Accessed: 13 May 2023].

[11] C. Glover, "PaperCut vulnerabilities exploited using LockBit and Cl0p ransomware – Microsoft." techmonitor.ai para. 1, Apr. 27, 2023. [Online]. Available: https://techmonitor.ai/technology/cybersecurity/papercut-vulnerability-lockbit-clop-microsoft-ransomware [Accessed: 14 May 2023].

[12] L. H. Newman, "Ransomware's Dangerous New Trick Is Double-Encrypting Your Data." wired.com para. 1, May 17, 2021. [Online]. Available: https://www.wired.com/story/ransomware-double-encryption/ [Accessed: 15 May 2023].

[13] R. Lakshmanan, "LockBit 3.0 Ransomware: Inside the Cyberthreat That's Costing Millions." thehackernews.com para. 6, Mar. 18, 2023. [Online]. Available: https://thehackernews.com/2023/03/lockbit-30-ransomware-inside.html [Accessed: 15 May 2023].

[14] T. Mclntosh, A.S.M. Kayes, Y.P. Chen, A. Ng, and P. Watters, (2023) "Applying staged event-driven access control to combat ransomware," Computers & Security, 128. doi: 10.1016/j.cose.2023.103160.

[15] L. Irwin, "Lack of education is the leading cause of successful ransomware attacks." itgovernance.co.uk para. 4, Oct. 24, 2019. [Online]. Available: https://www.itgovernance.co.uk/blog/lack-of-education-is-the-leading-cause-of-successful-ransomware-attacks [Accessed: 15 May 2023].

[16] S. Young, "When ransomware strikes, what's your recovery plan?" Network Security, 2021(7). doi: 10.1016/S1353-4858(21)00077-5.